\documentclass[dvips]{article}
\usepackage{icrctc07}

\title{REAS2:\\CORSIKA-based Monte Carlo simulations of geosynchrotron radio emission}
\shorttitle{REAS2 geosynchrotron simulations}
\authors{T. Huege$^{1}$, R. Ulrich$^{1}$, R. Engel$^{1}$.}
\shortauthors{T. Huege and et al}
\afiliations{$^1$Institut f\"ur Kernphysik, Forschungszentrum Karlsruhe, Postfach 3640, 76021 Karlsruhe, Germany}
\email{tim.huege@ik.fzk.de}

\abstract{Simulations of geosynchrotron radio emission from extensive air showers 
performed with the Monte Carlo code REAS1 used analytical parameterisations 
to describe the spatial, temporal, energy and angular particle distributions 
in air showers. The successor REAS2 replaces these parameterisations with 
precise, multi-dimensional histograms derived from per-shower CORSIKA 
simulations. REAS2 allows an independent selection between parameterisation 
and histogram for each of the relevant particle distributions, enabling us to 
study the changes arising from using a more realistic air shower model in 
detail. We describe the new simulation strategy and illustrate the effects 
introduced by the improved air shower model.}

\begin{document}
\maketitle

\section{Introduction}

From results of the LOPES experiment \cite{FalckeNature2005}, radio emission from cosmic ray air showers is known to be dominated by a geomagnetic emission mechanism that can be described with the ``geosynchrotron model'' \cite{FalckeGorham2003}. In the geosynchrotron process, relativistic secondary shower electrons and positrons are deflected in the earth's magnetic field, thereby giving rise to strongly pulsed, coherent radio emission in the frequency range from $\sim10$~to~100~MHz.
In the recent years, the geosynchrotron model had evolved from analytic frequency-domain calculations \cite{HuegeFalcke2003a} to time-domain Monte Carlo simulations based on an analytical description of the underlying extensive air shower, as implemented in the REAS1 simulation code \cite{HuegeFalcke2005a}. Many general properties of the radio emission have been predicted since using REAS1 \cite{HuegeFalcke2005b}.
The successor to this code, REAS2, now features an implementation of the geosynchrotron model no longer based on analytically parameterised air shower properties, but using realistic, per-shower CORSIKA \cite{HeckKnappCapdevielle1998} simulations to model the spatial, temporal, energy and angular distributions of the shower particles. In this article, we describe the changes arising from the much more realistic air shower model and illustrate the potential of the new simulation code for advanced studies of the radio pulse shape.

\section{REAS2 air shower model}

While the electromagnetic emission model has not changed between REAS1 and REAS2, REAS2 simulations are now based on very detailed particle information derived with CORSIKA on a per-shower basis. For each individual air shower, CORSIKA writes out separate information for electrons and positrons sampled in (usually) 50 layers between the point of first interaction and the observer position. Each layer encompasses
\begin{itemize}
\item{one three-dimensional histogram of}
  \begin{enumerate}
  \item{particle arrival time relative to that of an imaginary primary particle propagating with the speed of light from the point of first interaction}
  \item{lateral distance of the particle from the shower core}
  \item{particle energy}
  \end{enumerate}
\item{and one three-dimensional histogram of}
  \begin{enumerate}
  \item{angle of the particle momentum to the shower axis}
  \item{angle of the particle momentum to the (radial) outward direction}
  \item{particle energy.}
  \end{enumerate}
\end{itemize}
These histograms give REAS2 access to a true four-dimensional distribution of particles in atmospheric depth, arrival time, lateral distance and energy, and in addition describe the angular distribution of particle momenta as a function of particle energy and atmospheric depth. The chosen separation of the distributions into two histograms ensures that the necessary amounts of data can be handled on standard PCs while making approximations of only minor significance for the simulation of the radio signal. (The most important drawback of this scheme is the loss of information on azimuthal asymmetries in the air shower.) Naturally, effects associated with air showers induced by different types of primary particles can be analysed in detail with this simulation strategy. The longitudinal evolution of the air shower is sampled on an additional, finer grid (usually) spaced with 5~g~cm$^{-2}$ distance.

\section{REAS2 vs.\ REAS1 results}

\begin{figure*}[!htb]
\centering
\includegraphics[width=4.4cm,angle=270]{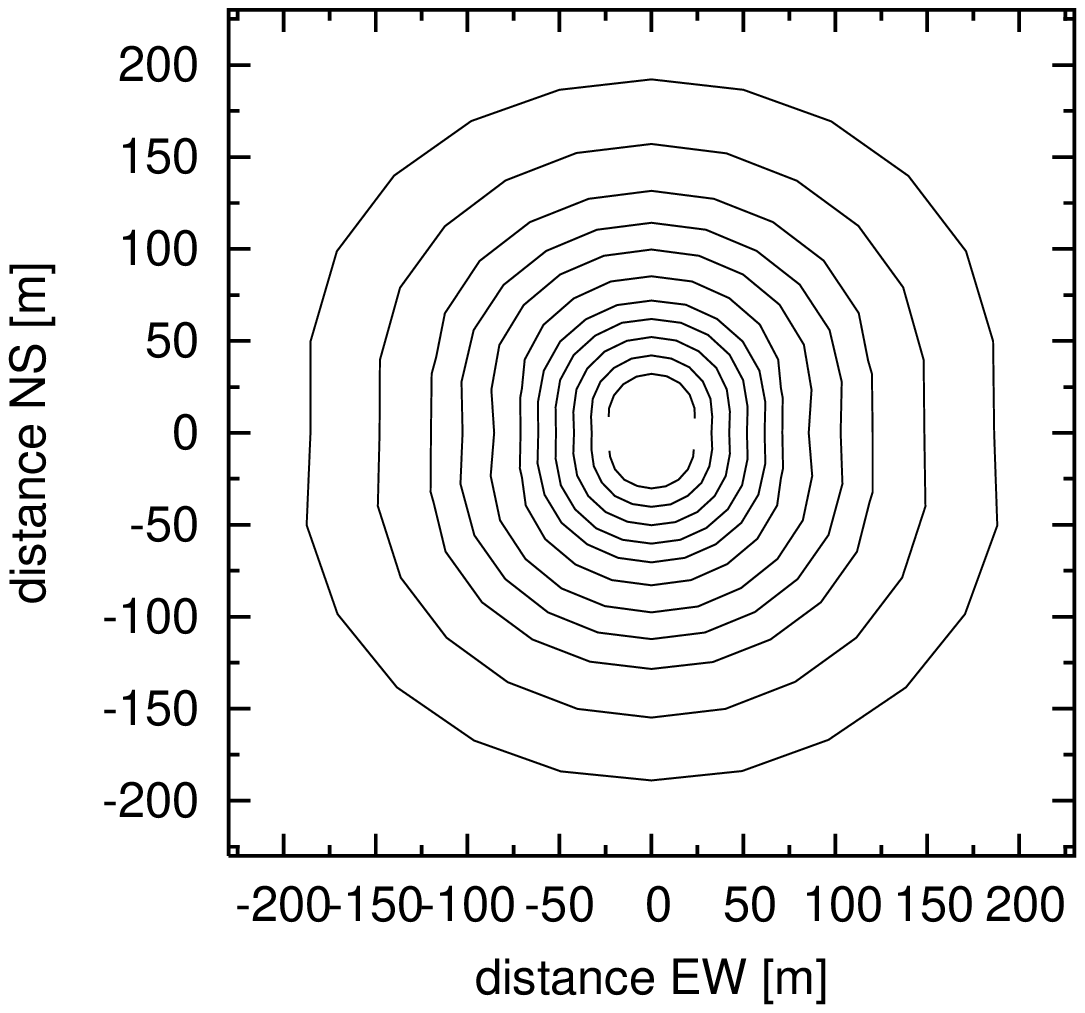}
\includegraphics[width=4.4cm,angle=270]{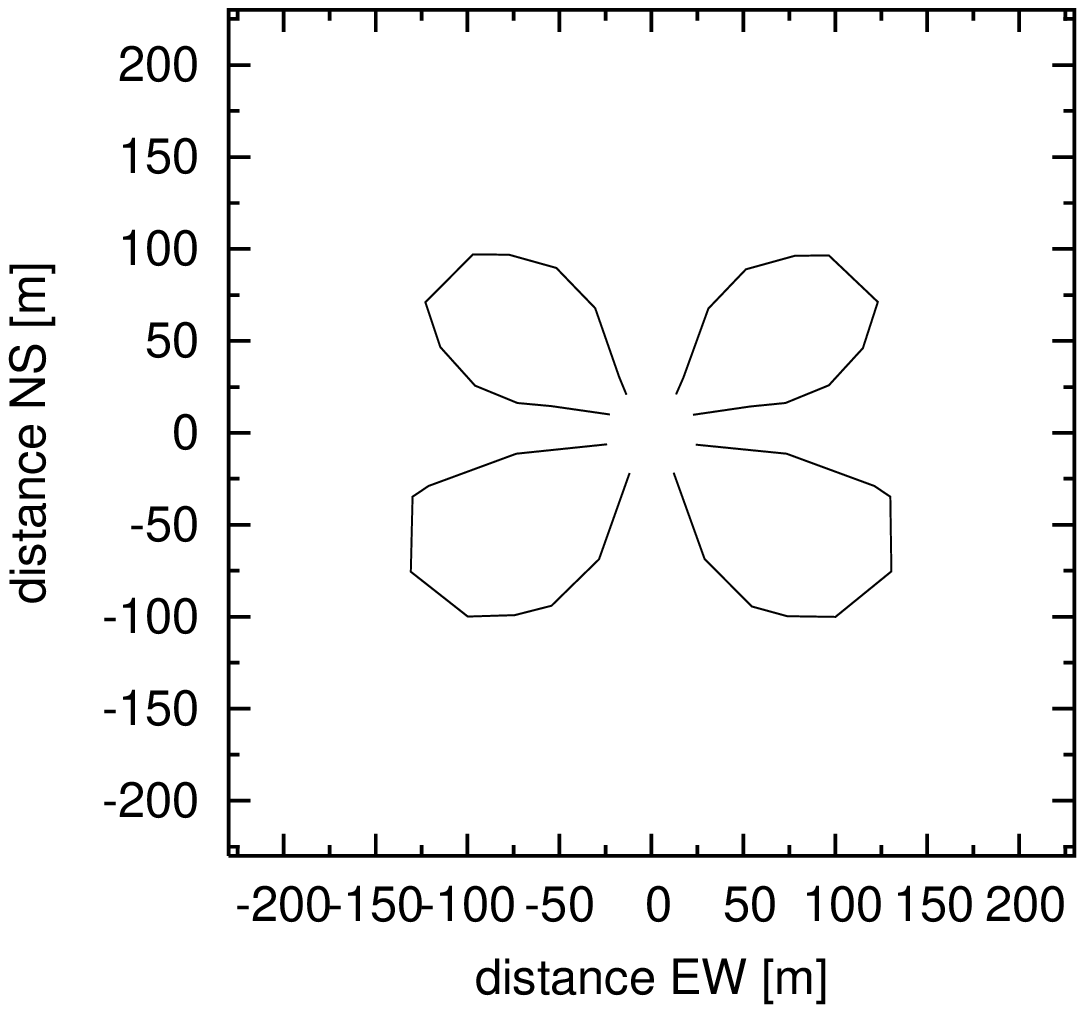}
\includegraphics[width=4.4cm,angle=270]{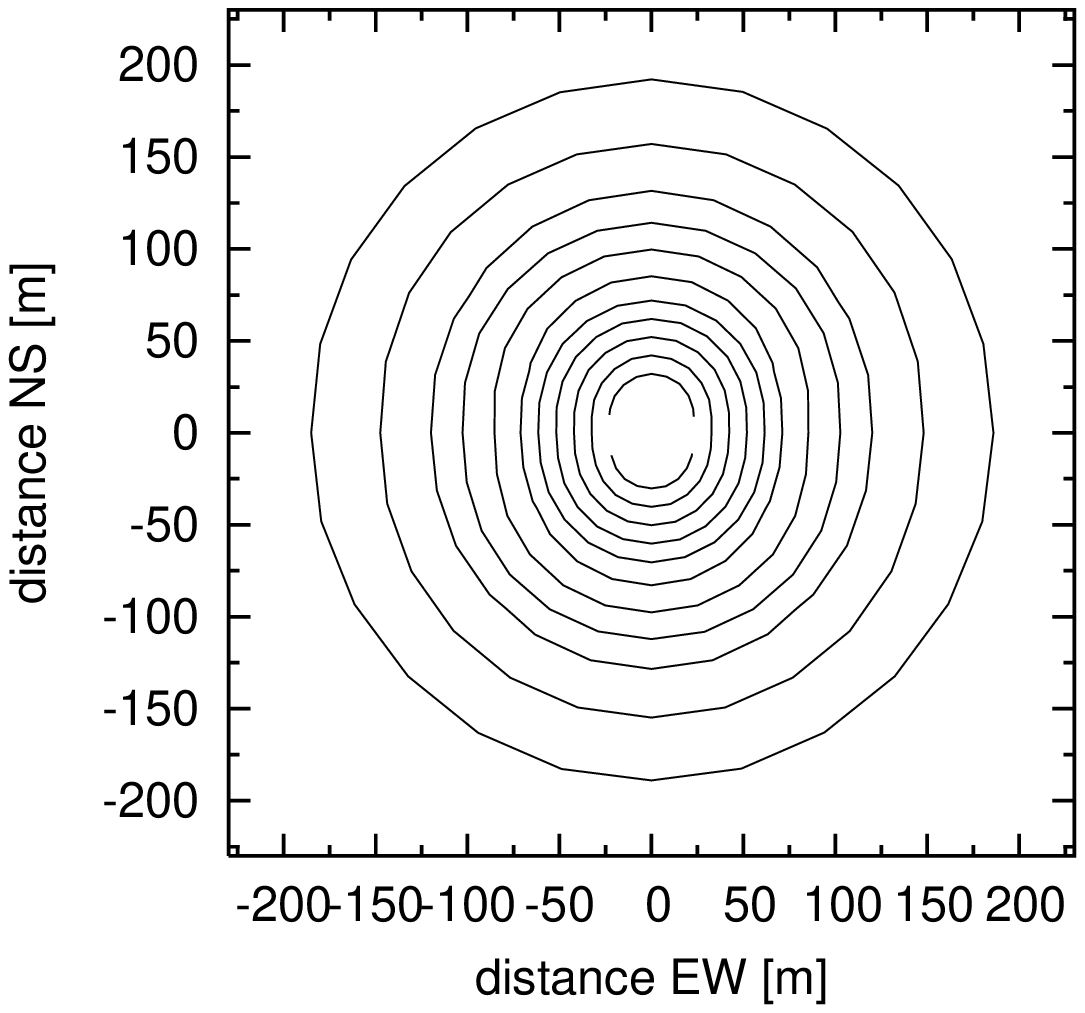}\\
\includegraphics[width=4.4cm,angle=270]{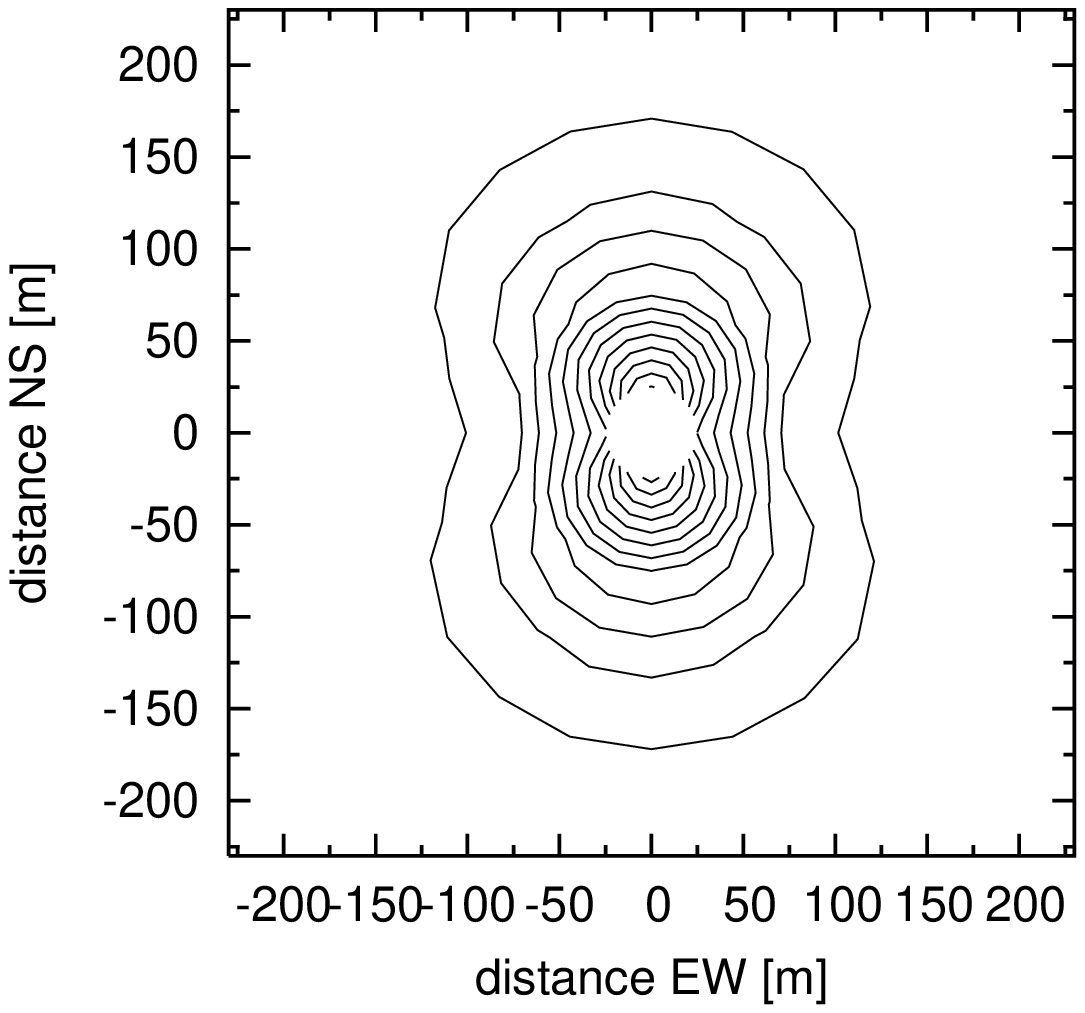}
\includegraphics[width=4.4cm,angle=270]{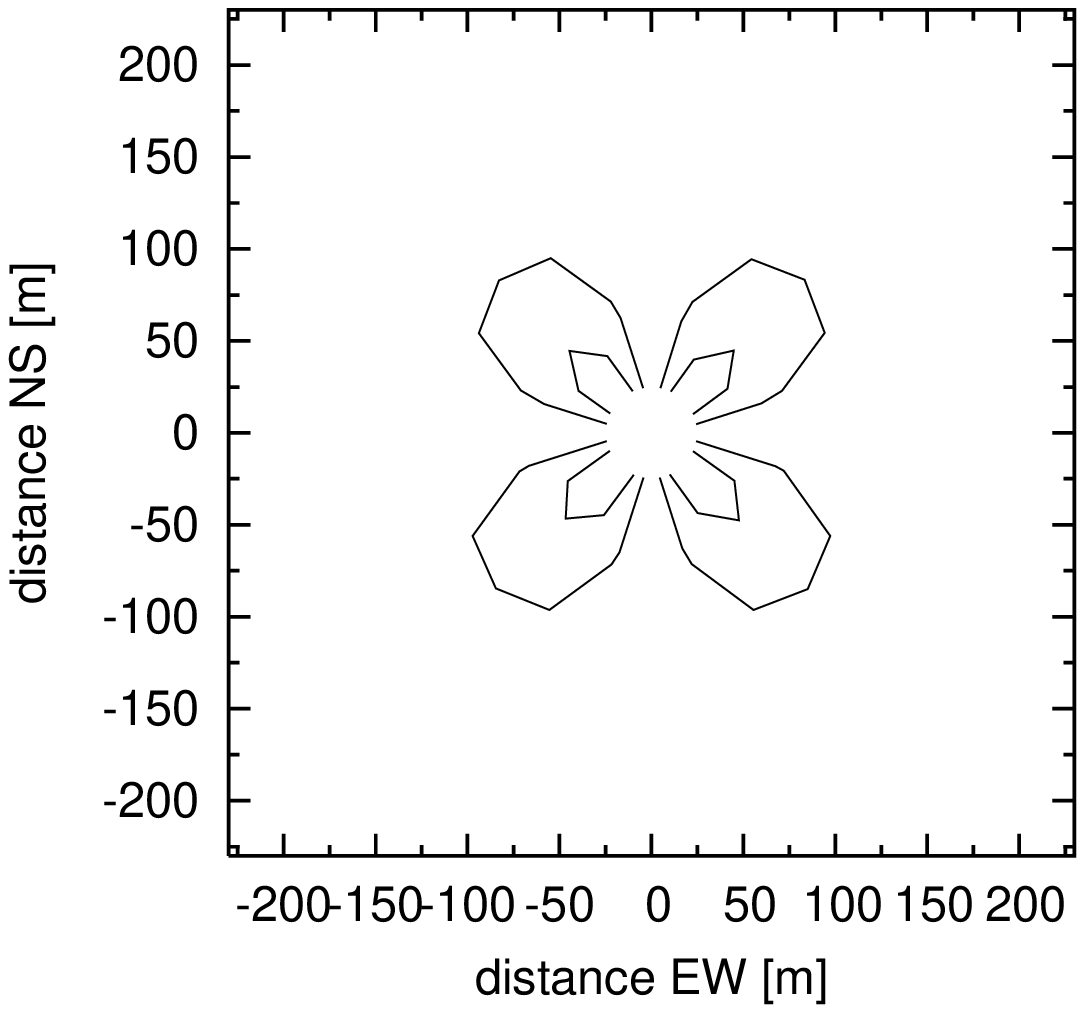}
\includegraphics[width=4.4cm,angle=270]{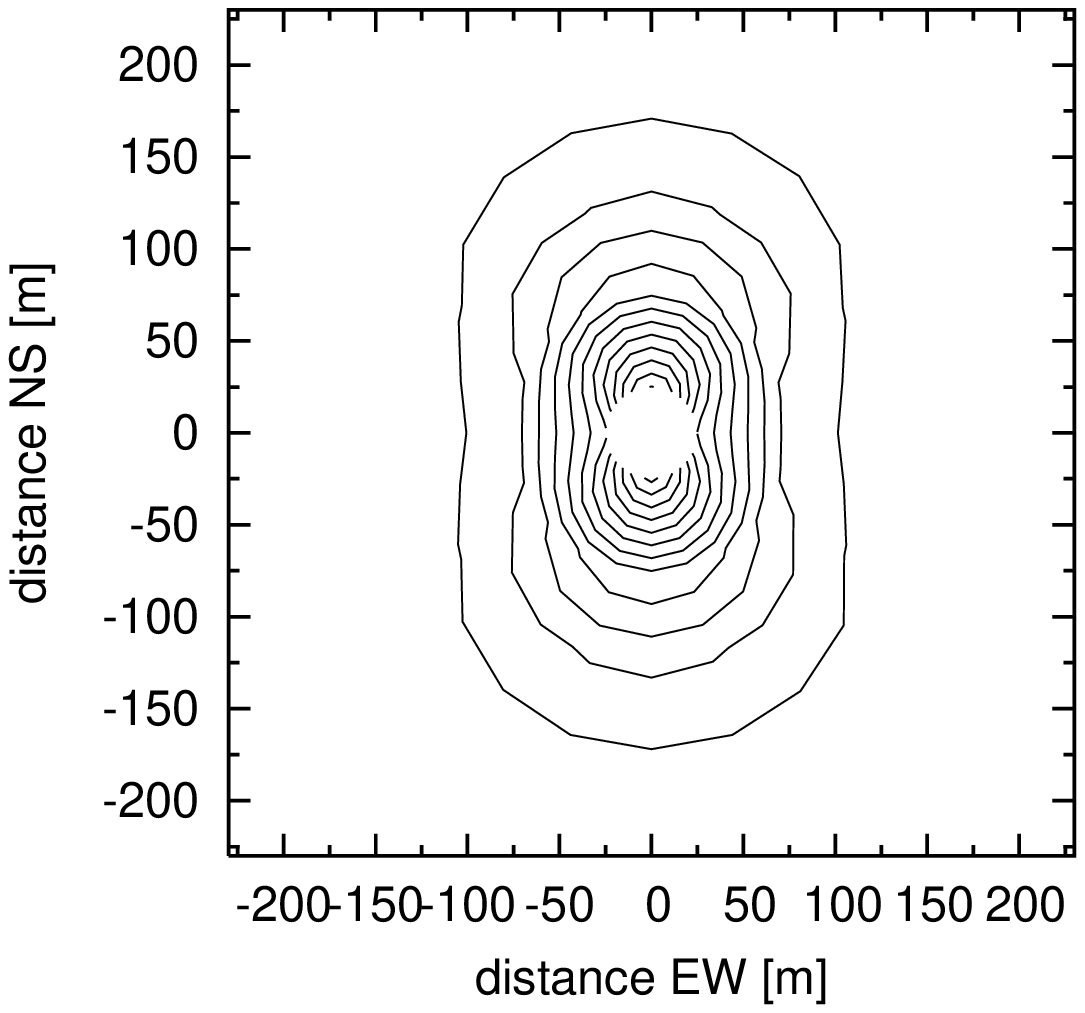}\\
\caption{Contour plots of REAS1 (upper) and REAS2 (lower) air shower emission at $\nu=60$~MHz. The columns (from left to right) show the total field strength, the north-south polarisation component and the east-west polarisation component. The vertical polarisation component (not shown here) does not contain any significant flux. Contour levels are 0.25~$\mu$V~m$^{-1}$~MHz$^{-1}$ apart in $E_{\omega}$, outermost contour corresponds to 0.25~$\mu$V~m$^{-1}$~MHz$^{-1}$. White centre region has not been calculated.
\label{fig:contours}}
\end{figure*}

Incorporating the new CORSIKA-based air shower model and some additional enhancements, the REAS2 code provides a much more realistic determination of geosynchrotron radio emission than its predecessor REAS1. A particular merit of the chosen approach is that the transition from the REAS1 to the REAS2 air shower model can be performed in a gradual fashion, switching the different particle distributions (spatial, temporal, energy, and angular) from parameterised to histogrammed one at a time and analysing the changes arising in the radio signal. The corresponding effects and technical details have been discussed elsewhere \cite{HuegeUlrichEngel2007a}; here we only compare the end result of REAS2 simulations with those of REAS1 simulations for the typical reference case of a vertical $10^{17}$~eV proton-induced air shower.
\begin{figure*}[!htb]
\begin{minipage}{0.47\textwidth}
\centering
\includegraphics [width=4.80cm,angle=270]{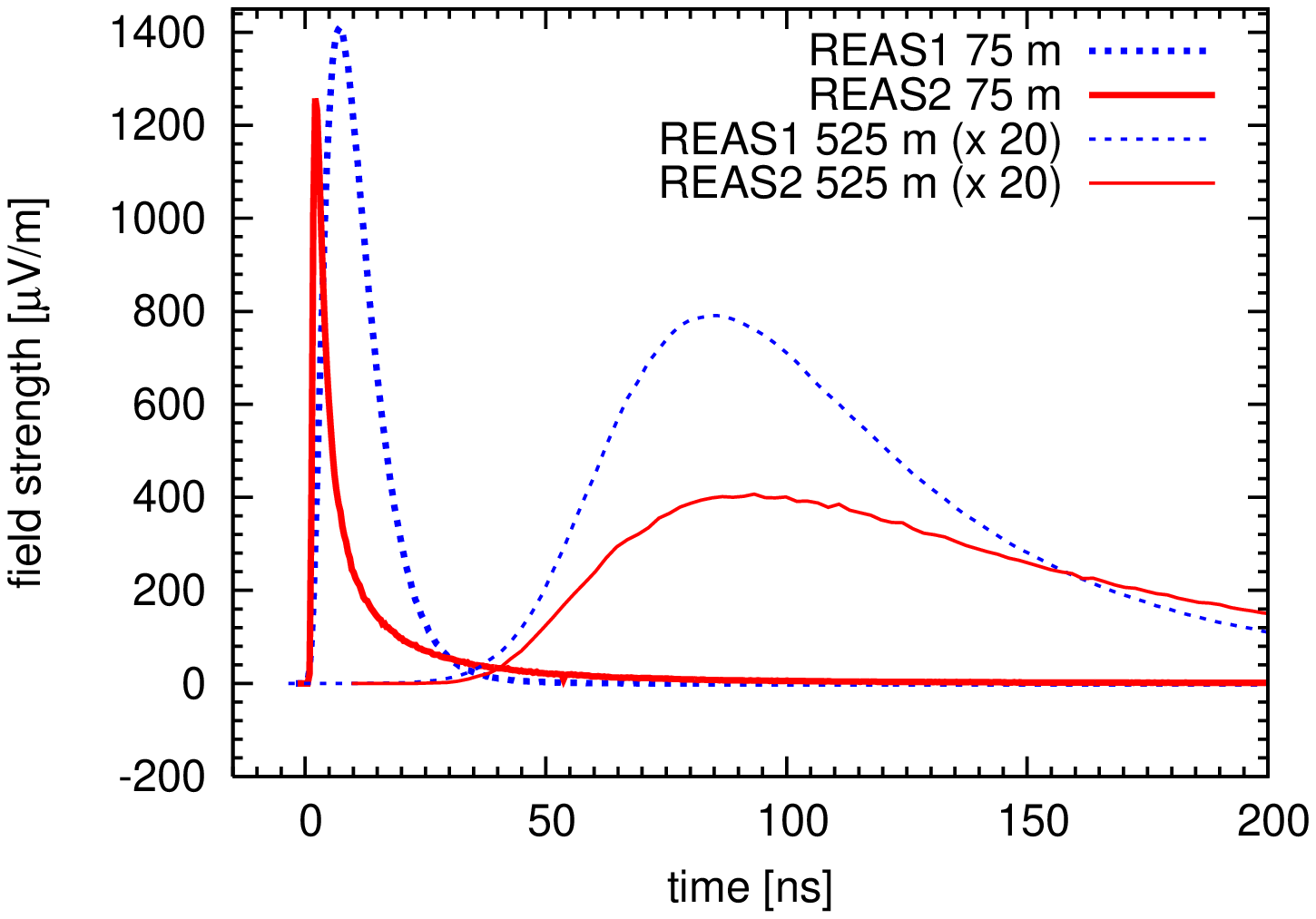}
\caption{Comparison of REAS1 and REAS2-simulated radio pulses at 75~m and 525~m north of the shower core.}\label{fig:pulsesnorth}
\end{minipage}
\hspace{0.05\textwidth}
\begin{minipage}{0.47\textwidth}
\centering
\includegraphics [width=4.80cm,angle=270]{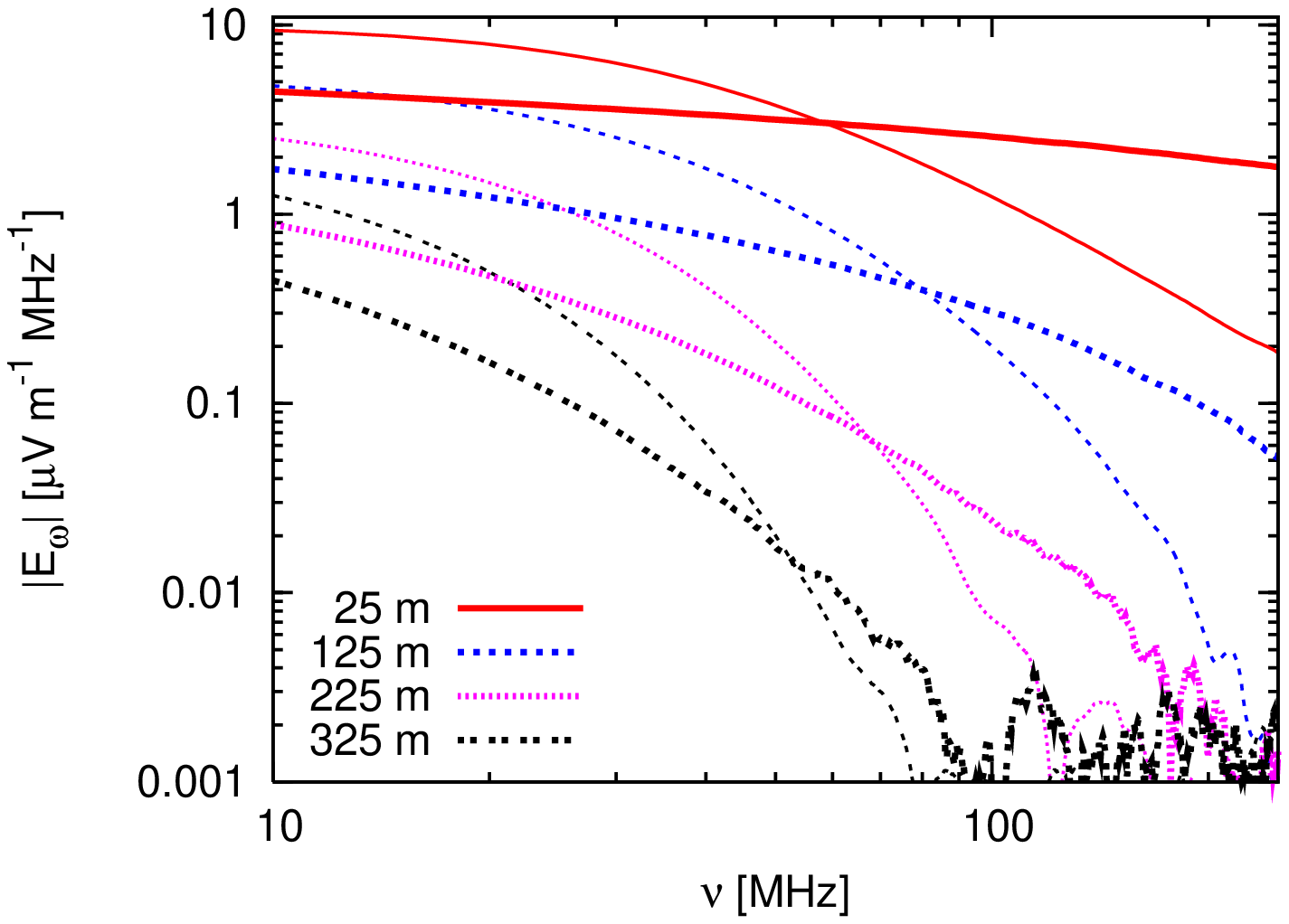}
\caption{Comparison of frequency spectra for the REAS1 (thin) and REAS2 (thick) simulated radio pulses.}\label{fig:spectranorth}
\end{minipage}
\end{figure*}
As presented in Fig.\ \ref{fig:pulsesnorth}, the radio pulses calculated with REAS2 for an observer to the north of the shower core show only moderate changes in comparison with the REAS1-generated pulses. In particular, the pulses close to the core (75~m corresponds to the typical lateral distance in the LOPES experiment) get considerably narrower, caused by the narrower arrival time distributions provided by CORSIKA in comparison with the parameterisation used in REAS1. Consequently, the frequency spectrum of the emission close to the core gets much flatter for the REAS2-calculated pulses (Fig.\ \ref{fig:spectranorth}). Further away from the core (525~m corresponds to the distance range of interest for larger scale radio antenna arrays), the amplitude drops by a factor of $\sim 2$, mainly as a consequence of the much broader angular distribution of particle momenta derived from CORSIKA. Interestingly, the overall field strength in the frequency band used by the LOPES experiment (40~to~80~MHz) does not change significantly between REAS1 and REAS2 (Fig.\ \ref{fig:spectranorth}).
\begin{figure}[!htb]
\centering
\includegraphics [width=4.80cm,angle=270]{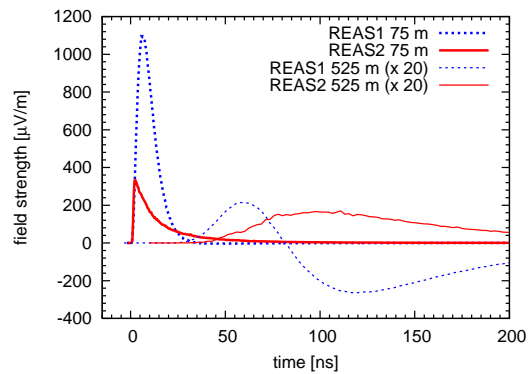}
\caption{Comparison of REAS1 and REAS2-simulated radio pulses at 75~m and 525~m west of the shower core. The bipolar pulses present in the REAS1 simulations are no longer present.}\label{fig:pulseswest}
\end{figure}
A more significant change becomes visible when studying the changes to the radio pulses for an observer west of the shower core, as depicted in Fig.\ \ref{fig:pulseswest}. First, a significant drop in pulse amplitude close to the shower core can be identified. This leads to a pronounced east-west versus north-south asymmetry in the radio ``footprint'', even for vertical air showers (cf.\ Fig.\ \ref{fig:contours}). At larger distances, a qualitative change in the pulse shape takes place: while REAS1-generated pulses in this region showed bipolar structures, the REAS2-calculated signals become universally unipolar. The bipolar pulse shapes in the REAS1 calculations can be considered artifacts of over-simplified particle distributions (e.g., in the momentum angles), and the REAS2 results describe the geosynchrotron emission much more realistically.
Another important characteristic of geosynchrotron radiation is its mostly linear polarisation. The comparison of the individual polarisation components depicted in Fig.\ \ref{fig:contours} demonstrates that the polarisation characteristics are identical between REAS1 and REAS2. At the same time, the contour plots illustrate once more the newly arising east-west versus north-south asymmetry and confirm that the absolute field strengths in the centre of the LOPES band (60~MHz) do not change considerably.

\section{Pulse shape analyses}

\begin{figure*}[!htb]
\begin{minipage}{0.47\textwidth}
\centering
\includegraphics [width=4.80cm,angle=270]{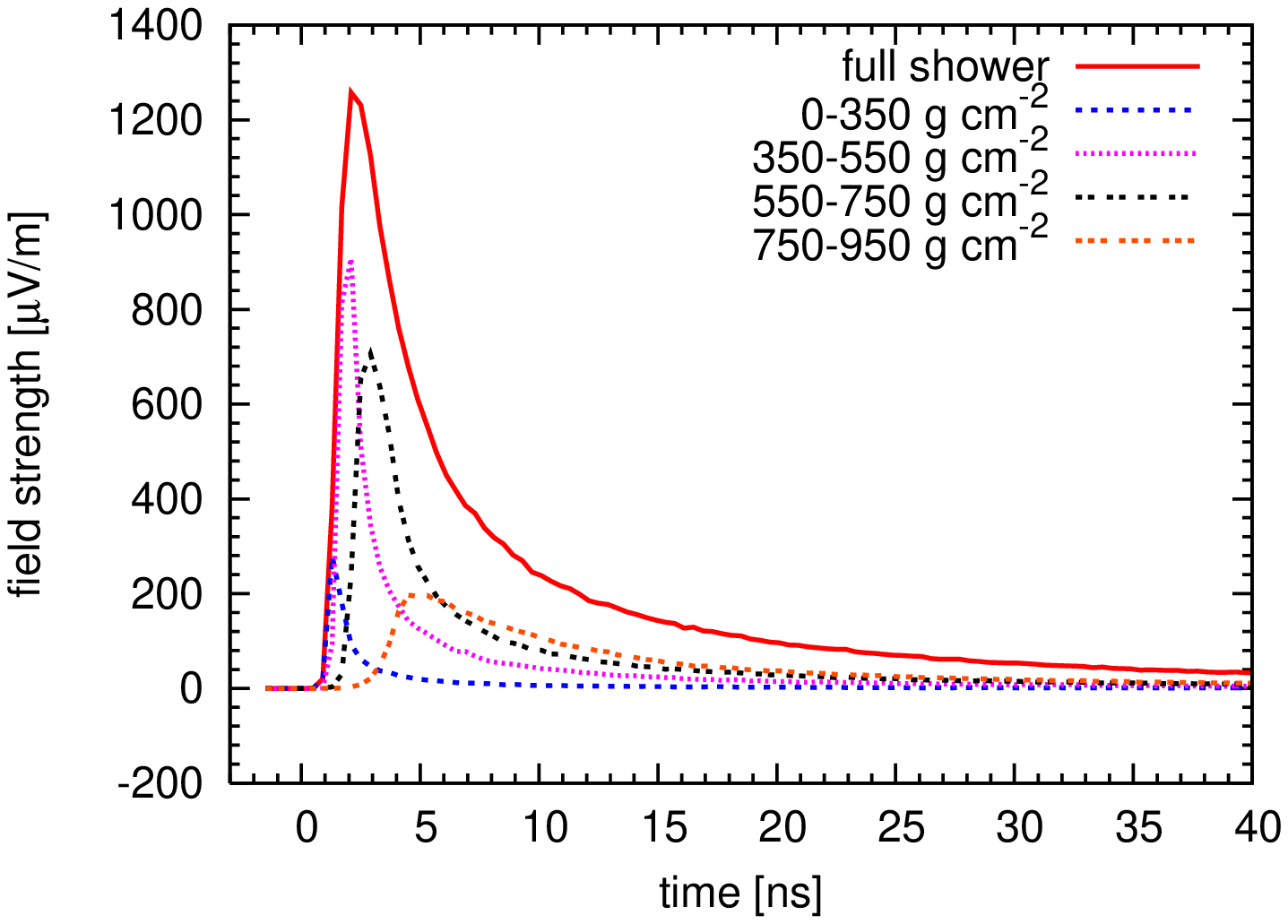}
\caption{Contribution of different longitudinal shower evolution stages to the radio pulse at 75~m north from the shower core.}\label{fig:depthregimes75m}
\end{minipage}
\hspace{0.05\textwidth}
\begin{minipage}{0.47\textwidth}
\centering
\includegraphics [width=4.80cm,angle=270]{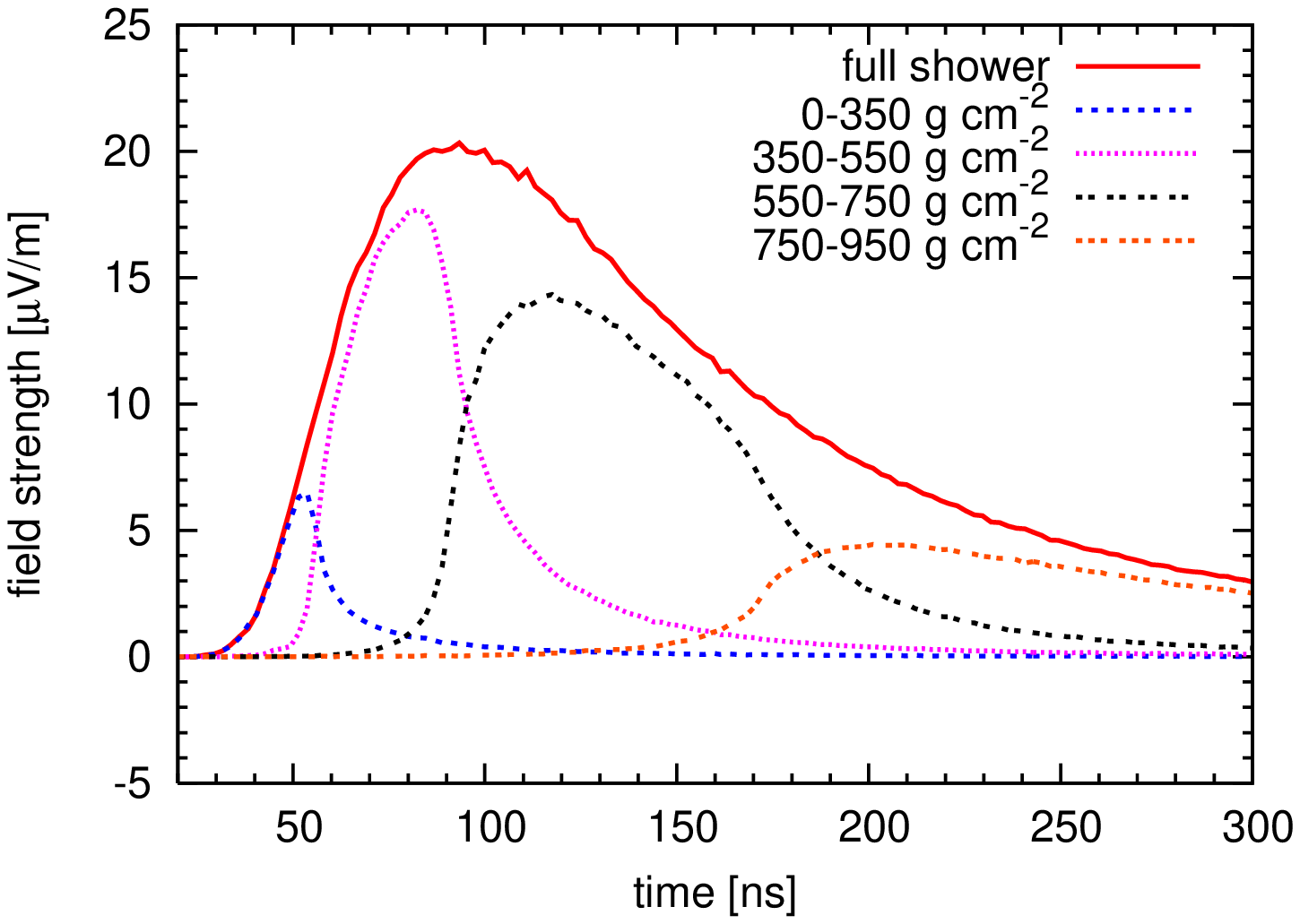}
\caption{Contribution of different longitudinal shower evolution stages to the radio pulse at 525~m north from the shower core.}\label{fig:depthregimes525m}
\end{minipage}
\end{figure*}

The highly detailed CORSIKA-based air shower model implemented in REAS2 allows advanced studies of the radio pulse shape. A particularly interesting question is how different phases in the longitudinal development of the air shower contribute to the radio pulses, as illustrated in Figs.\ \ref{fig:depthregimes75m} and \ref{fig:depthregimes525m}. Close to the shower core, signals from all over the longitudinal shower evolution arrive approximately simultaneously at the observer. The pulse shape close to the shower core thus gives a direct estimate of the overall particle arrival time distribution. (This could change once the refractive index profile of the atmosphere is taken into account.) At larger distances, geometrical time delays become important; the pulse shape thus provides direct information on the shower evolution profile. Another interesting result is that the emission is dominated by the shower maximum (here at 640~g~cm$^{-2}$) and the stage shortly before. The information content of the radio pulses can be exploited to estimate the primary particle energy and type from radio measurements on a shower-to-shower basis \cite{HuegeIcrc2007a}. Analyses how different particle energy ranges or radial distance ranges contribute to the radio signal have also been performed \cite{HuegeUlrichEngel2007a}.

\section{Conclusions}

With REAS2, a sophisticated Monte Carlo implementation of the geosynchrotron model based on a realistic, CORSIKA-based air shower model is now available. The transition from REAS1 to REAS2 has been carried out in a controlled way, and the changes arising are well-understood. In spite of the major model improvements, the changes are only moderate, in particular in the frequency range of current experiments. REAS2 can be used for in-depth studies of the information content of geosynchrotron radio pulses and as such is a powerful tool to unlock the full potential of the radio technique for cosmic ray measurements.


\end{document}